\documentclass{ws-procs975x65}

\usepackage{graphicx}
\usepackage{amsmath}
\usepackage{amsfonts}
\usepackage{amssymb}
\usepackage{amsbsy}

\newcommand{\Ham}{\mathcal{H}}
\newcommand{\Hil}{\mathcal{H}}
\newcommand{\ket}[1]{{|#1\rangle}}
\newcommand{\N}{\mathcal{N}}
\newcommand{\re}{\mathbb{R}}
\newcommand{\Pl}{{\rm Pl}}
\newcommand{\lPl}{\ell_{\Pl}}
\newcommand{\exps}[1]{{\langle {#1} \rangle}}

\newcommand{\rd}{{\rm d}}
\newcommand{\mbb}[1]{\boldsymbol{#1}}
\newcommand{\ub}[1]{\underline{#1}}
\newcommand{\id}{\mathbb{I}}
\newcommand{\expx}[1]{{\langle #1 \rangle}}

\begin{document}

\title{\uppercase{Dust time in quantum cosmology}}

\author{\uppercase{Viqar Husain}}

\address{Department of Mathematics and Statistics, University of New Brunswick,\\
Fredericton, NB  E3B 2S9, Canada \\
E-mail: vhusain@unb.ca}

\author{\uppercase{Tomasz Paw{\L}owski}}

\address{Department of Mathematical Methods in Physics, Faculty of Science, University of Warsaw\\
Ho\.za 74, 00-682 Warsaw, Poland\\
E-mail: tpawlow@fuw.edu.pl}

\begin{abstract} 
We give a formulation of quantum cosmology with a pressureless dust and arbitrary 
additional matter fields. The dust provides a natural time gauge corresponding to 
a cosmic time,  yielding a physical time independent Hamiltonian. 
The approach simplifies the analysis of both Wheeler-deWitt and loop quantum 
cosmology models, broadening the applicability of the latter. 
\end{abstract}


\vskip 1.0 cm 
\bodymatter\bigskip
 
\noindent{{\bf Introduction:}} In the process of constructing a consistent 
non-perturbative theory of quantum gravity the principal obstacle is the problem 
of time: due to the time-reparametrization invariance of general relativity, the 
physical Hamiltonian is replaced by a Hamiltonian 
constraint in the canonical formulation. Explicit time gauge fixing leads in general 
to non-unique and gauge-dependent theory.

A possible solution is  deparametrization, where a time variable is provided 
by suitable matter fields. This method however usually produces a Hamiltonian 
of the form of a square root -- which is often too complicated to apply to physical scenarios.

One completion of the gravity quantization procedure\cite{hp-lqg} avoids these problems: one couples gravity 
(and other matter) with a single timelike irrotational  dust field\cite{bk-dust}. For this system an application of a natural time gauge 
distinguished by the dust field, and the diffeomorphism invariant formalism of loop 
quantum gravity\cite{lqg-out}, results in a theory with a true Hamiltonian which is 
not a square root.  Quantum evolution is described  by a Schr\"odinger equation with time independent 
Hamiltonian.

Here we discuss an application\cite{hp-lqc} of this development in both the Wheeler-DeWitt 
(WDW) and loop quantum cosmology (LQC) quantization. To illustrate the useful properties 
of the approach we present  an example of the  flat isotropic universe with 
cosmological constant.

\noindent{{\bf The model:}}
%
The theory is given by the Einstein-Hilbert action with timelike 
dust field $T$ and  an arbitrary matter Lagrangian ${\cal L}_m$
\begin{equation}\label{eq:action}\begin{split}
    S &= \frac{1}{4G} \int \rd^4x \sqrt{-g} R
	- \int \rd^4x \sqrt{-g} {\cal L}_m 
     - \frac{1}{2} \int  \rd^4 x \sqrt{-g} M ( g^{ab}\partial_a T \partial_b T + 1)  . 
\end{split}\end{equation}
where $M$ is a Lagrange multiplier.

In the canonical formalism this action leads to a Hamiltonian constraint linear in the 
canonical momentum $p_T$ of $T$ and the usual spatial diffeomorphism constraint. 
The dust field admits a natural time gauge $T=t$ in which the true Hamiltonian is
\begin{equation}\label{eq:ham}
  \mbb{H} \equiv -p_T = \mbb{H}_G + \mbb{H}_m .
\end{equation}
Here $\mbb{H}_G$ and $\mbb{H}_m$ are respectively the gravitational and the matter 
part of the Hamiltonian constraint resulting from \eref{eq:action}. 

In the flat isotropic setting considered here the spacetime metric is of the form
\begin{equation}\label{eq:metric}
  \rd s^2 = -\rd t^2 + a^2(t) (\rd {\rm x}^2 + \rd {\rm y}^2 + \rd {\rm z}^2) ,
\end{equation}
where the unity of the lapse follows directly from equations of motion derived from 
\eref{eq:ham}. Thus $t$ reproduces exactly the cosmic time.

To present the formalism we further restrict our attention to the case $\mbb{H}_m=0$ 
and $\Lambda=0$. Now, the complete information about the system is captured in the 
canonical pair $(v,b)$, where $v=\alpha^{-1}a^3$ (with $\alpha\approx 1.35\ell_{\rm PL}^3$)
and $\{v,b\}=2$.

\noindent{{\bf Quantization:}} The classical system may be quantized using two distinct  
techniques. These are the geometrodynamic (or Wheeler-DeWitt approach) and  
and loop techniques (Loop Quantum Cosmology) \cite{aps-imp}.

\noindent{{\rm $\bullet$}} {\underbar{{\it WDW (Schr\"odinger) quantization:}}
Here one uses the standard Schr\"odinger representation. The canonical variables 
are promoted to operators acting on the dense domain in $L^2_s(\re,\rd v)$ (where 
$s$ denotes symmetric functions). The 
(symmetrically ordered) Hamiltonian \eqref{eq:ham} takes the form
\begin{equation}
  \hat{\mbb{H}}_G = \frac{3\pi G}{2\alpha} \sqrt{|\hat{v}} \hat{B}^2 \sqrt{|\hat{v}|}.
\end{equation}
so \eref{eq:ham} leads to a time-dependent Schr\"odinger equation
\begin{equation}\label{eq:Schroed}
  i \partial_t \Psi = \hat{\mbb{H}}_G \Psi
\end{equation}
The operator $\hat{\mbb{H}}_G$ {\it admits a $1$-parameter family of self-adjoint extensions.}
For each of them there exists an invertible map $\ub{P}_{\beta}$ to auxiliary Hilbert spaces. 
In a new variable $x=-1/b$, proportional to inverse Hubble parameter, the time evolution
of the physical state is represented by a simple formula 
\begin{equation}
\label{eq:WDW-state}
  [\ub{P}_{\beta}\Psi](x,t) = \int_{\re^+} \rd k\, \tilde{\Psi}(k)\,
  [\theta(x) e^{ikx} + \theta(-x) e^{i\beta} e^{ikx}] e^{i\omega_k t} , 
  \quad 
  \omega_k = 3\pi\ell_{\rm Pl}^2\alpha^{-1} k ,
\end{equation}
This formula describes a coherent plain wave packet propagating freely through 
configuration space with exception of the point $x=0$ (corresponding to a classical 
singularity), where it is rotated by an extension-dependent angle.
The interesting physical observables, like the volume $\hat{V} = |\hat{a}^3|$ when 
mapped to auxiliary spaces take simple analytic form, allowing to evaluate the 
trajectory. In particular
\begin{equation}\label{eq:WDW-v-evol}
  \exps{\hat{V}}(t) = V(t) = 2\alpha^{-1} \exps{\hat{k}}\  [3\pi\lPl^2(t-t_s)]^2 + 2\alpha\tilde{\sigma}_x^2 ,
 \ \ \ \hat{k} = k\id , 
\end{equation}
where $t_s$ has an interpretation as the big crunch/bang time. and $\tilde{\sigma}_x$ 
is related to the dispersion of $\hat{x}$ at $t_s$.

\noindent{{\rm $\bullet$}} \underbar{{\it LQC quantization:}} The physical Hilbert space
is $\Hil = L^2(\bar{\re},\rd\mu_B)$, where $\bar{\re}$ is the Bohr compactification of
the real line and $\rd\mu_B$ is the Haar measure on it. The Hamiltonian $\mbb{H}_G$ 
is now a difference operator in $v$
\begin{equation}\label{eq:HG-L0-1}
  \hat{\Ham}_{G}\ =\ -\frac{3\pi G}{8\alpha} \sqrt{|\hat{v}|}
    (\hat{\N}-\hat{\N}^{-1})^2 \sqrt{|\hat{v}|} \ , \quad
  \hat{\N}\ket{v}\ =\ \ket{v+1} \ ,
\end{equation}
and is {\it non-negative definite and essentially self-adjoint. }
By introducing a variable $x=-\cot(b)$ and repeating the procedure applied in WDW 
quantization we arrive to the analog of the Schr\"odinger eq. \eqref{eq:Schroed}
generating the quantum evolution
\begin{equation}\label{eq:L0-state}
  [P\Psi](x,t)\ =\ \int_{\re^+} \rd k\ \tilde{\Psi}(k) e^{-ikx} e^{i\omega_k t} \ , \quad
  \tilde{\Psi} \in L^2(\re^+,k\rd k) \ ,
\end{equation}
where $\omega_k$ is given by \eqref{eq:WDW-state} and $P$ is a mapping analogous 
to $\ub{P}_{\beta}$. This in turn leads to a quantum trajectory
(where the meaning of $t_s$ and $\tilde{\sigma}_x^2$ is the same as in \eref{eq:WDW-v-evol})
\begin{equation}\label{eq:v-traj}
  V(t)\ =\ 2\alpha^{-1} \exps{\hat{k}}\  [3\pi\lPl^2(t-t_s)]^2
  + 2\alpha\tilde{\sigma}_x^2 + 2 \alpha \exps{\hat{k}}  .
\end{equation}
The above description easily generalizes to the case $\Lambda\neq 0$, where the Hamiltonian 
remains self-adjoint and the state evolution is still given by \eref{eq:L0-state} with
the form of $x(b)$ distinct from $\Lambda=0$ case but still explicitly known. 

\noindent{{\bf Comparison:}} The above examples are an explicit illustration of 
singularity resolution in LQC and the lack of  it in WDW, in the following sense.

\noindent{$\bullet$} In WDW, evolving the state across the singularity
requires specification of additional boundary data (choice of an extension). Furthermore 
the quantum trajectory $V(t)$ approaches $V=0$ up to quantum variation.

\noindent{$\bullet$} In LQC the quantum evolution is unique and the minimal volume 
of the universe is well isolated from zero. This illustrates the dynamical resolution of a singularity through a big bounce \cite{aps-imp}.

\noindent{{\bf Application - modified Friedmann equation:}} In the LQC quantization
the Hubble and  gravitational energy density operators 
\begin{equation}
  \hat{H} = \pi\lPl^2\alpha^{-1} \sin(2\hat{b}) , \quad
  \hat{\rho}_G\ =\ -\rho_c\sin^2(\hat{b}) , \quad \rho_c\approx 0.41\rho_{{\rm Pl}} .
\end{equation}
are physical observables.
The precise relation between their expectation values takes the form 
(where $\sigma_{H}$, $\sigma_{\rho}$ are the dispersions of $\hat{H}$ and $\hat{\rho}_G$
respectively)
\begin{equation}\label{eq:Fried-eff}
  \expx{\hat{H}}^2\
  =\ \frac{8\pi G}{3} \expx{-\hat{\rho}_G} \left( 1 - \frac{\expx{-\hat{\rho}_G}}{\rho_c} \right)\
    -\ \left[ \frac{8\pi G}{3} \frac{\sigma_\rho^2}{\rho_c} + \sigma_H^2 \right] \ ,
\end{equation}
which is the \emph{exact} realization of the so called \emph{modified Friedmann equation} 
valid for all physical states and the arbitrary matter content.

\bibliographystyle{ws-procs975x65}
\bibliography{main}

\begin{thebibliography}{1}

\bibitem{hp-lqg}
V.~Husain and T.~Paw{\l}owski, {\em Phys.Rev.Lett.} {\bf 108}, p. 141301
  (2012).

\bibitem{bk-dust}
J.~Brown and K.~V. Kucha{\v{r}}, {\em Phys.Rev.} {\bf D51}, 5600 (1995).

\bibitem{lqg-out}
H.~Nicolai, K.~Peeters and M.~Zamaklar, {\em Class.Quant.Grav.} {\bf 22}, p.
  R193 (2005).

\bibitem{hp-lqc}
V.~Husain and T.~Paw{\l}owski, {\em Class.Quant.Grav.} {\bf 28}, p. 225014
  (2011).

\bibitem{aps-imp}
A.~Ashtekar, T.~Paw{\l}owski and P.~Singh, {\em Phys.Rev.} {\bf D74}, p. 084003
  (2006).

\end{thebibliography}

\end{document}